\documentclass[aps,prl,reprint,preprintnumbers,amsmath,amssymb,floatfix]{revtex4-2}

\usepackage{longtable}
\usepackage{morefloats}
\usepackage[dvips]{graphicx}
\usepackage{color}
\usepackage{epsfig,graphicx,amsfonts,amsbsy}
\usepackage{amsmath,amsfonts,amsthm,amssymb}
\usepackage{appendix}
\usepackage{bbm}
\usepackage{makeidx}
\usepackage{url}
\usepackage{verbatim}
\usepackage[bookmarksnumbered,pdfpagelabels=true,plainpages=false,colorlinks=true,linkcolor=blue,citecolor=blue,urlcolor=blue]{hyperref}
\usepackage[rightcaption]{sidecap}
\usepackage{array}
\usepackage{booktabs}
\usepackage{multirow}
\usepackage{bbm}
\usepackage{tabularx}
\usepackage{cancel,soul,ulem}

\newcommand{\bs}[1]{\boldsymbol{#1}}

\begin{document}
\title{Bosonic Bott Index and Disorder-Induced Topological Transitions of Magnons}
\author{X. S. Wang}
\author{Arne Brataas}
\author{Roberto E. Troncoso}
\email{r.troncoso@ntnu.no}
\affiliation{Center for Quantum Spintronics, Department of Physics, Norwegian University of Science and Technology, NO-7491 Trondheim, Norway}

\begin{abstract}
We investigate the role of disorder on the various topological magnonic phases present in deformed honeycomb ferromagnets. To this end, we introduce a bosonic Bott index to characterize the topology of magnon spectra in finite, disordered systems. The consistency between the Bott index and Chern number is numerically established in the clean limit. We demonstrate that topologically protected magnon edge states are robust to moderate disorder and, as anticipated, localized in the strong regime. We predict a disorder-driven topological phase transition, a magnonic analog of the ``topological Anderson insulator" in electronic systems, where the disorder is responsible for the emergence of the nontrivial topology. Combining the results for the Bott index and transport properties, we show that bulk-boundary correspondence holds for disordered topological magnons. Our results open the door for research on topological magnonics as well as other bosonic excitations in finite and disordered systems.
\end{abstract}
\maketitle

%%%%%%%%%%%%%%%%%%%%%%%%%
%%INTRODUCTION
%%%%%%%%%%%%%%%%%%%%%%%%%
Topological phases in nature have attracted intense interest and research in recent decades. Initially, the research was mainly on electronic systems such as topological insulators \cite{TI1,TI2}. Currently, the concept of topology has been extended to many research areas that span condensed-matter physics. Different kinds of bosonic low-energy excitations, e.g., phonons \cite{TPhonon1}, photons \cite{TP1}, magnons
\cite{TM1,TM2,TM3,TM41,TM42,TM5,TM6,TM7,TM8,TM9,TM10,TM11,TM12,TM13,TM14,TM15,TM16},
and even macroscopic motions \cite{TCM1,PNAS,GaoPRB2020} host topological states. Topological phases are characterized by certain topological indices that remain unchanged under smooth deformations. Nontrivial topology is usually associated with the appearance of robust edge states immune to disorder, known as ``bulk-boundary correspondence", which is one of the most exotic features of topological matters and invokes many potential applications \cite{Stern1179,Yue2017,Zeng2020,Wang2017,Wang2018}.

Among various kinds of excitations, research on topological states in magnonic systems has increased in recent years. Many models have been proposed to support topological magnons. Some of them can be mapped to known electronic models \cite{TM3,TM5,TM7,TM12,TM13}, while some are exclusive in bosonic systems \cite{TM14,TM15,Wang2017,Wang2018}. Nevertheless, most previous studies on topological magnons
focused on clean systems and did not consider disorder, which is ubiquitous and unavoidable in nature. The role of disorder in topological systems is a crucial issue since it is related to one of the fundamental features of topological systems: the robustness of the edge states. In electronic systems research, there is plenty of discussion on this issue. Not only have transport properties been studied \cite{TAI1,TAI2,TAI3,Su2016},
but also the Chern number in real space \cite{Kitaev2006,Prodan2009}, and the Bott index \cite{Bott1,Bott2,Bott3,Bott4,
Bott5} has been used to label the topology of finite or disordered systems. However, how disorders affect topological magnons is still underexplored.

In this letter, we consider a honeycomb ferromagnet with nearest-neighbor (NN) pseudodipolar interaction \cite{pseudo},
whose magnons can be topologically nontrivial, and we study the effect of disordered on-site anisotropy.
To label the magnon topology in finite or disordered magnets, we generalize the real-space Bott index
\cite{Bott1,Bott2,Bott3,Bott4,Bott5} to bosonic systems.
We first show that the Bott index agrees with the Chern number in the clean limit.
We then demonstrate the bulk-boundary correspondence in disordered magnonic systems by comparing
their transport properties and topological indices. We find that for the topologically nontrivial phase,
the topology as well as the protected edge states are quite robust against disorder, unless the
disorder is more than 3 times larger than the gap. For the topologically trivial phase, we identify a disorder-induced
nontrivial phase, which is the magnonic analogy of the ``topological Anderson insulator" \cite{TAI1,TAI2}.
Our findings reveal that the Bott index is a useful tool in research on topological bosonic systems without
translational symmetry.

%%%%%%%%%%%%%%%%%%%%%%%%%
%%MODEL
%%%%%%%%%%%%%%%%%%%%%%%%%
\begin{figure*}
  \centering
  \includegraphics[width=0.9\textwidth]{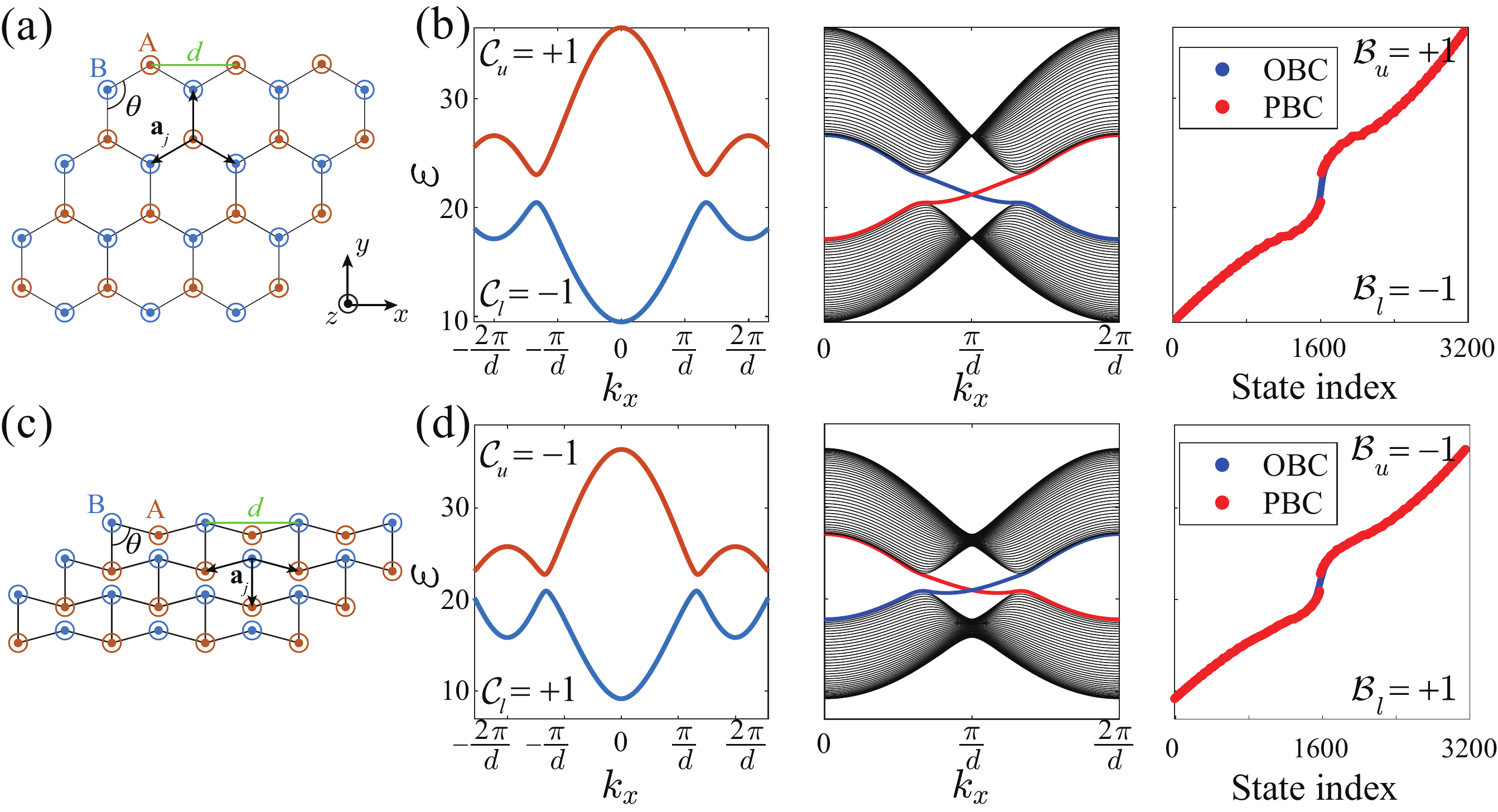}\\
\caption{(a) Schematic of a normal honeycomb magnet. The ground state is out-of-plane.
(b) From left to right: the spin-wave spectra for infinite samples (along $k_x$ for $k_y=0$), zigzag strips of width $N_y=100$ along $x$, and finite samples of $N_x=40$ and $N_y=40$.
(c) Schematic of a squeezed honeycomb magnet of $\theta=5\pi/12$. (d) Spin-wave spectra for (c). }\label{band}
\end{figure*}

We consider a ferromagnetic material with localized spins on a
deformed two-dimensional honeycomb lattice formed by heavy metal atoms having strong spin-orbit coupling.
The spin Hamiltonian we consider is given by
\begin{multline}
\mathcal{H}=-J\sum_{\langle i,j\rangle}\mathbf{S}_i\cdot\mathbf{S}_j-F\sum_{\langle i,j\rangle}
\left(\mathbf{S}_i\cdot\mathbf{e}_{ij}\right)\left(\mathbf{S}_j\cdot\mathbf{e}_{ij}\right)\\
-\frac{1}{2}\sum_{i}K_iS_{zi}^2
-\mu_B B \sum_{i}S_{zi},
\label{hami}
\end{multline}
where $J>0$ is the NN ferromagnetic exchange coupling and $F$ is the NN pseudodipolar interaction resulting from the spin-orbit coupling, with $\mathbf{e}_{ij}$ being the unit vector connecting lattice sites $i$ and $j$ along one of the NN lattice vectors $\mathbf{a}_{1,2,3}$ \cite{pseudo}.
This Hamiltonian can be mapped to the ``Kitaev model" \cite{Kitaev2006} in linear regime. The easy-axis anisotropy $K_i$ at the A(B)-sublattice consists of two parts: a homogeneous part $K_\text{A(B)}$ and a random part $K_{ri}$, where $K_{ri}$ is statistically independent for different $i$ and uniformly distributed on the interval $[-W,W]$.
For convenience, we define $K=(K_\text{A}+K_\text{B})/2$ and $\Delta K=(K_\text{A}-K_\text{B})/2$.
$B$ is the applied magnetic field along the $\hat{\bs z}$-direction ($\mu_B$ is the Bohr magneton).
Since we are interested in the strong disorder limit, a sufficiently large easy-axis anisotropy and/or a magnetic field is assumed so that the spins align out-of-plane in the ground state \cite{Wang2018}.
The angles between $\mathbf{a}_{1,2,3}$ and the $x$-direction are $\theta_1=\pi/2$, $\theta_2=\pi/2-\theta$ and $\theta_3=\pi/2+\theta$. Figures \ref{band}(a) and (c) show a perfect ($\theta=2\pi/3$) and deformed ($\theta=5\pi/12$) honeycomb lattice, respectively.

We first consider the magnon spectra on honeycomb ferromagnets in the absence of disorder. The linear-excitation $\mathbf{k}$-space magnon Hamiltonian is written in terms of
bosonic creation/annihilation operators, $a^\dagger$($a$) and $b^\dagger$($b$) on each sublattice A and B
\cite{HolsteinPR1940} as
$\mathcal{H}=\frac{1}{2}x^\dagger H_\mathbf{k} x$, where $x=
\left(a_\mathbf{k},a_{-\mathbf{k}}^\dagger,b_\mathbf{k},b_{-\mathbf{k}}^\dagger\right)^T$
(see the Supplemental Materials \cite{SM} for the explicit form of $H_\mathbf{k}$).
We diagonalize $H_\mathbf{k}$ to obtain the magnon spectrum by employing the Bogoliubov transformation \cite{Bogoljubov1958}. The transformation matrix $\mathcal{T}_\mathbf{k}$ which diagonalizes $H_\mathbf{k}$ satisfies the
generalized eigenvalue problem (GEP) \cite{White1965}
\begin{equation}
\eta H_\mathbf{k}\mathcal{T}_\mathbf{k}=\mathcal{T}_\mathbf{k}\eta E_\mathbf{k},
\label{geneig}
\end{equation}
where $\eta$ is a metric matrix reflecting the commutation relations $\eta_{ij}=\left[x_i,x^\dagger_j\right]$ so that
$\eta=\mathbbm{1}_{2\times2}\otimes\sigma_z$ ($\sigma_{x,y,z}$ are the Pauli matrices), and $E_\mathbf{k}$ is the diagonal
matrix whose diagonal elements $\varepsilon_n(\mathbf{k})$ are the eigenvalues of $\mathcal{H}$.
The GEP has a particle-hole
symmetry in that the $\varepsilon_n$s are artificially doubled in positive-negative pairs $\varepsilon_n(\mathbf{k})=-\varepsilon_n(\mathbf{-k})$.
Therefore, it is sufficient to consider the positive solutions of $\varepsilon_n$ only.
Note that Eq.~\eqref{geneig} is equivalent to the result from the linearized classical Landau-Lifshitz-Gilbert equation
\cite{Wang2017}. This system is known to be gapped and topologically nontrivial when $\theta\neq \frac{\pi}{2}$ \cite{Wang2017,PNAS}. In Fig.~\ref{band}, we show the spin-wave spectra of three different samples for the normal (a) and deformed (c) honeycomb lattices. From left to right, in (b) and
(d), the spectrum is plotted for the infinite system (along $k_x$ with $k_y=0$), the zigzag strip along the $x$-direction and width $N_y=100$, and the finite samples with dimensions $N_x=N_y=20$, assuming periodic boundary conditions (PBCs) and open boundary conditions (OBCs). The parameters are $F=7J$, $K=20J$, $\Delta K=0$, $B=0$, and $N_x$ and $N_y$ are the number of units cells in the $x$- and $y$-directions, respectively ($K=20J$ and $B=0$ are used throughout this letter).
Both (b) and (d) show gapped bulk
spectra in infinite and periodic systems and gapless (crossing) edge states for an
open strip, indicating a nontrivial topology.

In infinite translationally symmetric systems, the Chern number of the $n^{\text{th}}$ band is \cite{TM14,TM15}
\begin{equation}
\mathcal{C}_n=\frac{1}{2\pi}\mathrm{Im}\int_{\mathrm{B.Z.}}\mathrm{tr}\left[P_n\left(\frac{\partial P_n}{\partial k_x}\frac{\partial P_n}{\partial k_y}-
\frac{\partial P_n}{\partial k_y}\frac{\partial P_n}{\partial k_x}\right)\right]d\mathbf{k},
\end{equation}
Here, $P_n(\mathbf{k})$ is the bosonic projector defined by $P_n=\mathcal{T}_\mathbf{k}\eta\Gamma_n \mathcal{T}_\mathbf{k}^\dagger\eta$,
where $\Gamma_n$ is a diagonal matrix taking a value of $1$ for the $n^{\text{th}}$ diagonal
components and zero otherwise. The Chern numbers of the upper (lower) magnon bands, $\mathcal{C}_u$ ($\mathcal{C}_l$), are labeled in Figs. \ref{band}(b) and (d). {A topological transition occurs at $\theta=\pi/2$, where $\mathcal{C}_u$ and $\mathcal{C}_l$ flip their signs.} When $\theta \gtrless\frac{\pi}{2}$, $\mathcal{C}_u=-\mathcal{C}_l=\pm 1$  \cite{PNAS}. When the magnetic anisotropy at each sublattice differs; i.e., $\Delta K\neq 0$, one of the gaps at K or K$^\prime$ points closes and reopens, and the system becomes topologically trivial ($\mathcal{C}_u=\mathcal{C}_l=0$) \cite{Wang2017,Wang2018}.

In the presence of disorder or in finite samples, the periodicity of the lattice
is broken so that the $\mathbf{k}$-space Chern number is invalid. We need a real-space index to label the topology.
In electronic systems, the Bott index was introduced to study nonperiodic systems such as
disordered topological insulators \cite{Bott3} and quasicrystals \cite{Bott4,Bott5}.
The Bott index quantifies the obstruction to construct an orthogonal basis of localized Wannier functions that span the occupied states \cite{Bott2}, and it has been proven
to be equivalent to the Chern number in the large-system limit
\cite{Toniolo2017}.

We now extend the definition of the Bott index to bosonic systems. For a finite system of size $N_x\times N_y$ (in total, there are $N=N_xN_y$ unit cells), dual to the $\mathbf{k}$-space representation, the GEP in real space is $\eta H \mathcal{T}=\mathcal{T}\eta E$, where $H$ is the $4N\times4N$ real-space Hamiltonian, $\eta=\mathbbm{1}_{2N\times 2N}\otimes \sigma_z$ is the metric due to the bosonic commutation relation in real space, and $E$ is the diagonal matrix of eigenenergies. $\mathcal{T}$ is the matrix diagonalizing the Hamiltonian. For a set of
eigenstates $\left\{\varepsilon_n\right\}$, its {\it bosonic Bott index} is given by
\begin{equation}
\mathcal{B}\left\{\varepsilon_n\right\}=\frac{1}{2\pi}\mathrm{Im}\left\{\mathrm{tr}[\log(VUV^\dagger U^\dagger)]\right\},
\end{equation}
where the two matrices $U$ and $V$ are defined from
\begin{gather}
Pe^{2\pi i X}P=\mathcal{T}\eta \left(\begin{matrix}0 & 0 \\0 & U \end{matrix}\right)\mathcal{T}^\dagger\eta,\\ Pe^{2\pi i Y}P=\mathcal{T}\eta \left(\begin{matrix}0 & 0 \\0 & V \end{matrix}\right)\mathcal{T}^\dagger\eta,
\end{gather}
where $P=\mathcal{T}\eta \Gamma \mathcal{T}^\dagger\eta$ is the projector on states $\left\{\varepsilon_n\right\}$.
$X=i_x/N_x$ and $Y= i_y/N_y$ are the rescaled coordinates,
where $i_{x,y}$ are spatial indices of the unit cells. $\Gamma$ is a diagonal matrix taking a value of $1$ for the $j^\text{th}$ diagonal elements when $j\in\left\{\varepsilon_n\right\}$, and 0 otherwise. Note that for fermionic systems, $\eta=\mathbbm{1}$, and the above definition returns to the electronic Bott index \cite{Toniolo2017,Bott5}. $\mathcal{B}$ is always an integer as long as $VUV^\dagger U^\dagger$ is nonsingular \cite{Bott3,Toniolo2017}, and specifically, $\mathcal{B}=0$ when the matrices $U$ and $V$ commute. For a clean system with well-defined gaps, the Bott index of each band separated by gaps is well defined.

\begin{figure}
  \centering
  \includegraphics[width=8.5cm]{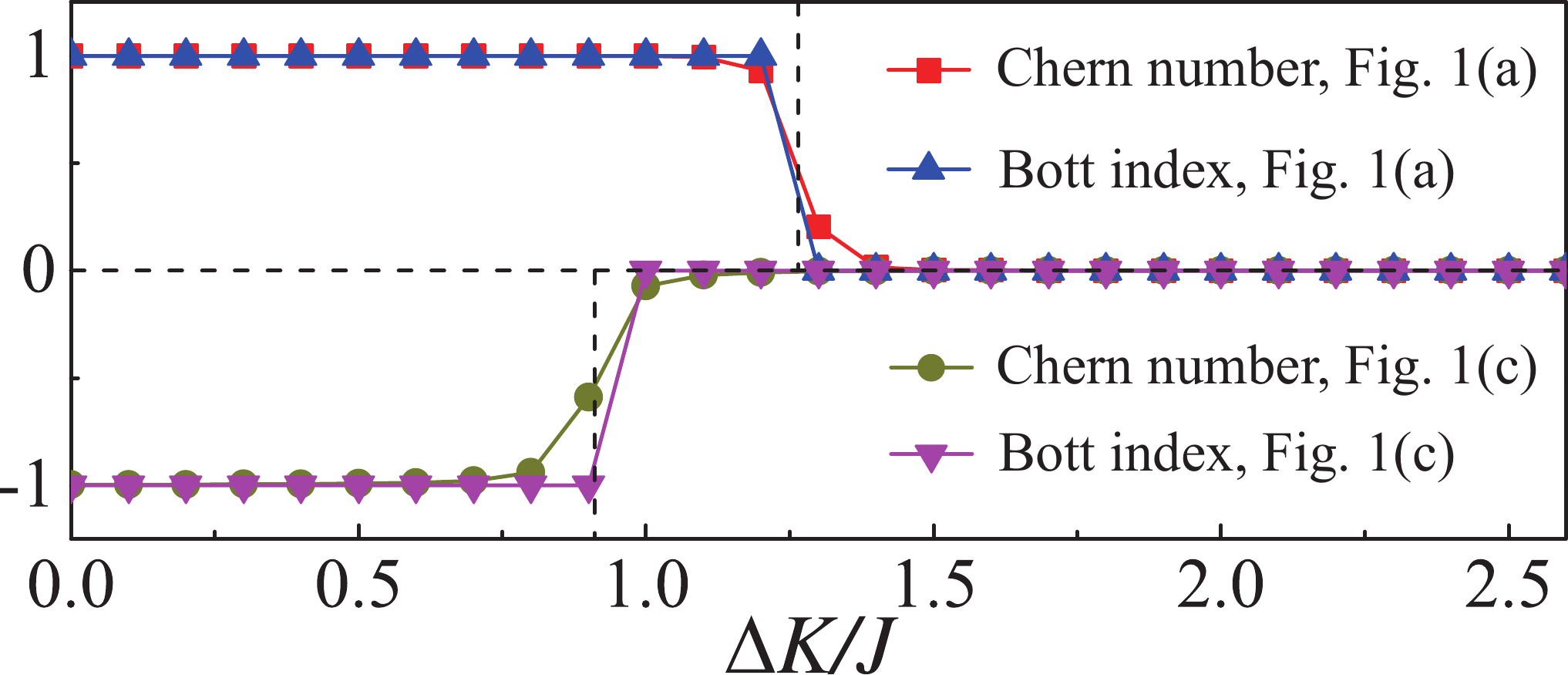}\\
\caption{Comparison between the Chern number ($\mathcal{C}_u$) and Bott index ($\mathcal{B}_u$) for finite and clean systems. The equivalence is established as a function of the staggered anisotropy $\Delta K$ and for a system size of $40\times 40$.}
  \label{compare}
\end{figure}

We then compared the Bott index and Chern number in the absence of disorder ($W=0$). In Figs.~\ref{band}(b) and (d) (third panel), we label the Bott indices for the upper and lower magnon bands ($\mathcal{B}_u$ and $\mathcal{B}_l$, respectively) of clean $40\times 40$ samples with PBCs. The results are consistent with the Chern number for infinite systems.
A systematic comparison is shown in Fig.~\ref{compare} in terms of $\Delta K$ for the upper band of the normal [Fig.~\ref{band}(a)] and deformed [Fig.~\ref{band}(c)] honeycomb lattices. The vertical dashed line represents the $\Delta K$ values for
gap-closing, resulting in a topological phase transition from nontrivial to a trivial magnon spectrum. Both the Bott index and Chern number consistently describe the topology of the system. Note that near the topological transition point, the Berry curvature is ill defined, so the numerically calculated Chern numbers are not integers. Although the Bott indices are still integers in the case, a larger system size and thus higher computational cost are necessary to obtain accurate results.

Next, let us consider the presence of disorder in the conventional honeycomb lattice system [Fig.~\ref{band}(a)].
Because of disorder, the gap is filled with states even though PBCs are used. We define the
Bott indices as functions of energy \cite{RN553}, $\mathcal{B}_u(\varepsilon)$ and $\mathcal{B}_l(\varepsilon)$, where
$\mathcal{B}_u(\varepsilon)$ ($\mathcal{B}_l(\varepsilon)$) is the Bott index of all the states with higher (lower) energy than
$\varepsilon$. In the following, we calculate the ensemble-averaged Bott indices over 100 uncorrelated random configurations,
denoted by $\bar{\mathcal{B}}$, for system size of $N_x=N_y=40$.

\begin{figure}[!]
  \centering
  \includegraphics[width=8.5cm]{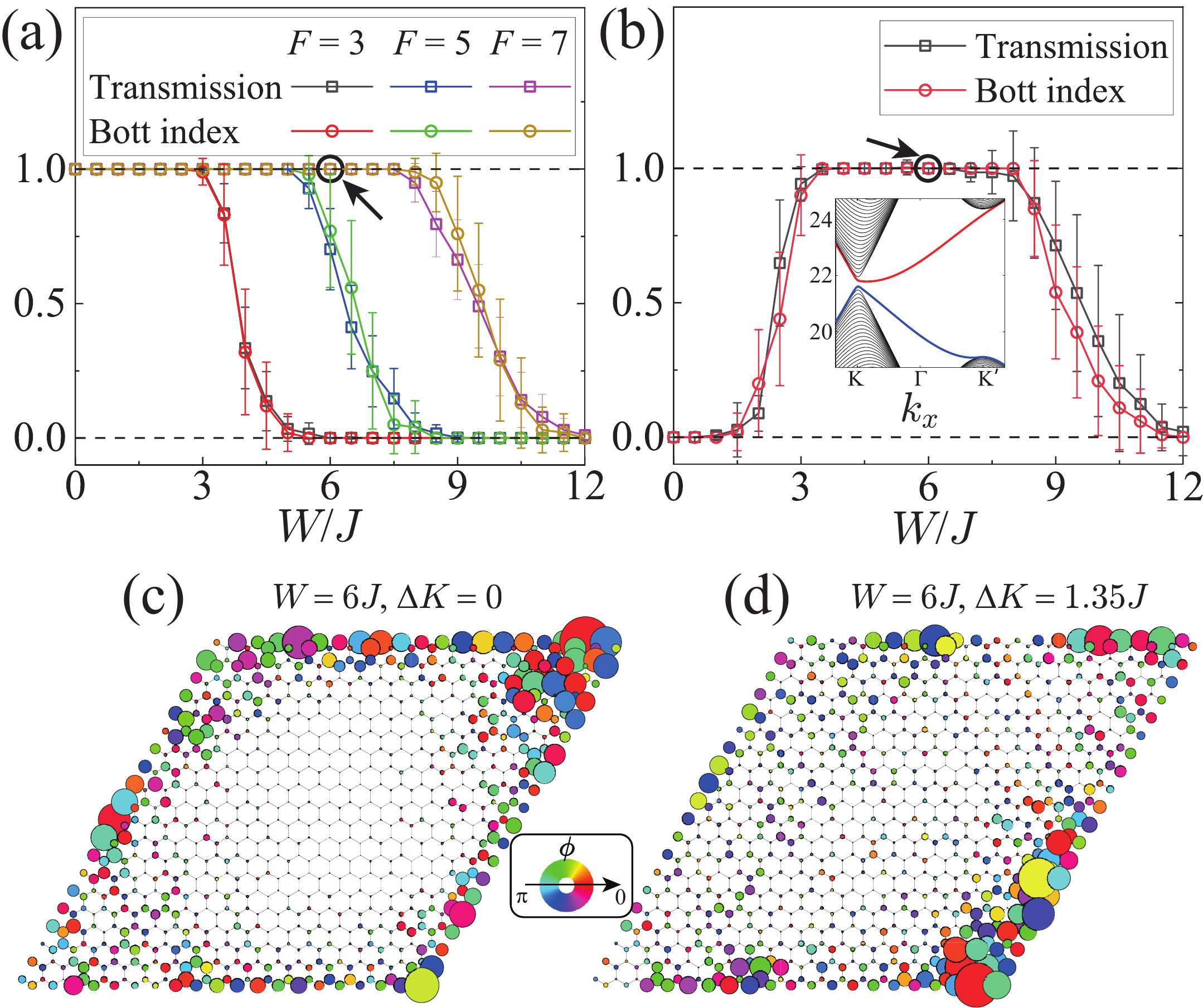}\\
\caption{Comparison between the Bott index and the total transmission as a function of the
disorder strength $W/J$, when the clean-limit system is (a) topologically nontrivial ($\Delta K=0$) and (b) trivial ($\Delta K=1.35J$). The inset in (b) is the band structure near the gap of a 100-wide zigzag strip. The real-space wave function is depicted for $F=7J$ and $W=6J$ in (c) $\Delta K=0$ and (d) $\Delta K=1.35J$, corresponding to the circled data points in (a) and (b), respectively. This is represented by the expectation of the in-plane spin components, $\langle{S_x}\rangle$ and $\langle{S_y}\rangle$, the spatial distribution of the eigenstate whose energy is closest to $\varepsilon_0$. The size of the circles indicates the amplitude, and the color encodes the azimuthal angle.
}\label{wf}
\end{figure}

First, we consider systems that are topologically nontrivial in clean limits. In Fig.~\ref{wf}(a), $\Bar{\mathcal{B}}_u(\varepsilon_0)$ is plotted against the disorder strength $W$ for $\Delta K=0$, $F=3J$, $5J$ and $7J$, where $\varepsilon_0$ is the energy at the midpoint of the gap in the clean limit \cite{SM}. For moderate disorder, the Bott index is still 1, meaning that the system is topologically nontrivial. When the disorder is strong enough, a topological transition occurs, and the system becomes topologically trivial. This phenomenon is consistent with the common wisdom that the topology is quite robust since very strong disorder (approximately 3 times the gap) is needed to break the topology. A more remarkable phenomenon occurs when the disorder affects a topologically trivial system. In Fig.~\ref{wf}(b), we consider an originally trivial system  (at $W=0$) with $F=7J$ and $\Delta K=1.35J$ and plot $\bar{\mathcal{B}}_u(\varepsilon_0)$ against $W$. The band structure of a strip near the gap in the clean limit is shown in the inset. There are no gapless edge states inside the bulk gap. Surprisingly, as the disorder strength increases, the Bott index increases from 0 and reaches a plateau of $\mathcal{B}_u(\varepsilon_0)=1$ and then drops to 0 at $W>8J$. This finding indicates that there exists a disorder-induced topological phase, similar to the ``topological Anderson insulator" phase in electronic systems \cite{TAI1,TAI2,TAI3}.

Now, we demonstrate the bulk-boundary correspondence in our topological magnon model by studying the transport properties.
We consider a disordered strip sample with two identical (clean) leads attached to its left- and right-hand sides.
We evaluate the total transmission probability $T(\varepsilon=\varepsilon_0)$ from left to right for $N_x=N_y=200$ samples \cite{SM}. The results are plotted in Figs.~\ref{wf}(a) and (b) (averaged over $100$ disorder realizations). In the clean limit, the total transmission equals the total number of propagating channels according to the Landauer-B\"{u}ttiker formula \cite{LB1988}. For the nontrivial phase, since there is one rightward edge channel, at zero disorder, we have $T=1$, as shown in Fig.~\ref{wf}(a). The topologically protected edge channel remains robust as the disorder increases; however, at certain values, the magnonic modes become localized, and thus, the topology is destroyed. For the trivial phase [Fig.~\ref{wf}(b)], since there is no channel inside the gap, we shift the band of the leads upward by $2J$ to make full use of the bulk channels \cite{TAI3}. As $W$ increases, the transmission
increases from 0 to a plateau ($T=1$) and then decays to zero at very large disorder, following the topological transition \cite{Martin2019}.

The existence of edge states in strongly disordered magnets is further confirmed by the calculation of the real-space wave functions. Eigenstates whose energies are closest to
$\varepsilon_0$ for a certain disorder configuration were considered. For clarity of representation, we use a smaller system size $N_x=N_y=20$. We plot the expectations of the in-plane spin components, $\langle{S_x}\rangle$ and $\langle{S_y}\rangle$, in Figs.~\ref{wf}(c) and (d) for the originally nontrivial phase ($\Delta K=0$) and disorder-induced nontrivial phase ($\Delta K=1.35J$), respectively. The parameters $F=7J$ and $W=6J$ were used for both plots; see the circled data points in Fig.~\ref{wf}(a)(b). Clear features of the edge states can be observed. However, for $\Delta K=1.35J$, the penetration depth is larger, so interedge backscattering is more likely; see the Supplemental Materials \cite{SM} for details.

To further understand the emergence of the disorder-driven topological transition, we consider the self-energy, $\Sigma$, induced
by the disorder, defined by $(\varepsilon_0-H_{\mathbf{k}}-\Sigma)^{-1}=\langle(\varepsilon_0-H_{\mathbf{k}}^\text{eff})^{-1}\rangle$,
where $H_{\mathbf{k}}^\text{eff}$ is the disorder-renormalized effective Hamiltonian. We numerically calculate $\Sigma$ in the self-consistent Born approximation
\cite{TAI2} for the parameters used in Fig.~\ref{wf}(b). The result is a $4\times 4$ matrix that can be decomposed into
three Hermitian components $\Sigma^{0\sim 2}$ and one anti-Hermitian component.
$\Sigma^0$ is proportional to identity matrix $\mathbbm{1}_{4\times4}$, which shifts the whole spectrum.
$\Sigma^1$ is proportional to $\sigma_x\otimes\mathbbm{1}_{2\times2}$, which shifts only the position of the valley.
$\Sigma^2$ is proportional to $\sigma_z\otimes \mathbbm{1}_{2\times2}$, which has the same structure as the $\Delta K$ term
in $H_\mathbf{k}$ and is responsible for the topological transition. I
The randomness on the anisotropy effectively reduces $\Delta K$, and drives the system back to the
non-trivial phase. The non-Hermitian component
reflects the inverse lifetime of the magnon states. By letting $\Sigma^2$ be the critical value of the topological transition, we can solve for the critical disorder strength $W=3.7J$, which is consistent with the numerical result \cite{SM}.

We have numerically demonstrated very good agreement between the total transmission and the Bott index,
which indicates that the bulk-boundary correspondence holds in the disordered topological magnon system.
Our numerical studies pave the way for a rigorous mathematical proof of the equivalence
between the bosonic Bott index and Chern number \cite{Toniolo2017}, as well as the bulk-boundary correspondence
in topological magnonic systems, which are open issues for future research.

All the discussions above also apply for other deformed honeycomb lattices, provided that the clean system is gapped \cite{PNAS}. Note that a further increase in $W$ destroys the ferromagnetic ground state, and the system enters a spin-glass-like state \cite{Edwards1975IOP}, which is not the purpose of this study. We expect our definition of the bosonic Bott index to be applied to any bosonic system \cite{SM}, as long as the metric matrix $\eta$ is modified according to the commutation relations of creation/annihilation operators, which will benefit many research areas, such as topological phononics, photonics and superconductors.
AB$_3$-type 2D honeycomb magnetic materials such as CrI$_3$ and OsCl$_3$
are possible experimental platforms for our model. There are already first-principles and  experimental indications of strong pseudodipolar
interaction \cite{OsCl3,CrI3}.

%%%%%%%%%%%%%%%%%%%%%%%%%
%%CONCLUSION
%%%%%%%%%%%%%%%%%%%%%%%%%
In conclusion, we introduced a bosonic Bott index as an integer-valued real-space topological invariant in bosonic systems and used it to study the magnon topology in a disordered honeycomb ferromagnet. In the clean limit, the topological phase is controlled by the bond angle and staggered anisotropy, and the Bott index is consistent with the Chern number. In the presence of disorder, the edge states in the nontrivial phase are robust to moderate disorder. In the trivial phase, the disorder can induce a phase transition to a nontrivial topology, which is the magnonic counterpart of the ``topological Anderson insulator" phase in electronic systems. Our findings open the door for the investigation of the topology of disordered bosonic systems.

\begin{acknowledgments}
This work was partially supported by the Research Council of Norway through its Centres of Excellence funding scheme, project no. 262633, ``QuSpin". X.S.W. acknowledges support from the Natural Science Foundation of China (grant no. 11804045).
\end{acknowledgments}

\clearpage
\onecolumngrid
\section{Supplemental Materials}
\setcounter{table}{0}
\setcounter{equation}{0}
\setcounter{figure}{0}
\renewcommand\theequation{S\arabic{equation}}
\renewcommand*{\citenumfont}[1]{S#1}
\renewcommand*{\bibnumfmt}[1]{S#1}
\renewcommand\thefigure{S\arabic{figure}}

\subsection{Some details of the magnon spectra}

The explicit form of matrix $H_\mathbf{k}$ is
\begin{equation}
H_\mathbf{k}=\left(\begin{matrix}
M_\mathrm{A} & 0 & -f(\mathbf{k}) & g_+(\mathbf{k}) \\
0 & M_\mathrm{A} & g_-(\mathbf{k}) & -f(\mathbf{k}) \\
-f^*(\mathbf{k}) & g_-^*(\mathbf{k}) & M_\mathrm{B} & 0 \\
g_+^*(\mathbf{k}) & -f^*(\mathbf{k}) & 0 & M_\mathrm{B} \\
\end{matrix}\right),
\end{equation}
where $M_\alpha=K_\alpha+\mu_BB+3J$ ($\alpha\in \text{A,B}$), $f(\mathbf{k})=\left(J+\frac{F}{2}\right)\sum_j e^{i\mathbf{k}\cdot \mathbf{a}_j}$ and $g_\pm(\mathbf{k})=\frac{F}{2}\sigma_j e^{\pm2i\theta_j}e^{i\mathbf{k}\cdot \mathbf{a}_j}$.
$H_\mathbf{k}$ is Hermitian.
The generalized eigenvalue problem $\eta H_\mathbf{k}\mathcal{T}_\mathbf{k}=\mathcal{T}_\mathbf{k}\eta E_\mathbf{k}$
is equivalent to an ordinary eigenvalue problem $H^\prime_\mathbf{k}\mathcal{T}_\mathbf{k}=\mathcal{T}_\mathbf{k}E^\prime_\mathbf{k}$,
where $H^\prime_\mathbf{k}=\eta H_\mathbf{k}$ is a non-Hermitian matrix and $E^\prime_\mathbf{k}=\eta E_\mathbf{k}$.
The non-Hermitian eigenvalue problem has real spectra when the ferromagnetic  $0^{\text{th}}$-order state is an energy-local-minimum state.

In the absence of disorder, for a perfect honeycomb lattice, the energies of the two bands at the K (K$^\prime$) point
are $M+(-)\Delta K$ and $\sqrt{M^2-\frac{9}{4}F^2}-(+)\Delta K$ ($M=K+\mu_BB+3J$). The midgap energy
$\varepsilon_0=\frac{1}{2}\left(M+\sqrt{M^2-\frac{9}{4}F^2}\right)$, independent of $\Delta K$.
The topological transition occurs at $\pm2\Delta K=M-\sqrt{M^2-\frac{9}{4}F^2}$, where
one of the gaps at the K and K$^\prime$ points closes.

For a deformed honeycomb lattice, due to the lack of 3-fold rotational symmetry, the Brillouin zone corners are no longer energy minima or maxima. The transition point and the midgap energy are calculated numerically.
\subsection{Transmission for different system sizes}
\begin{figure}[h]
  \centering
  \includegraphics[width=0.9\textwidth]{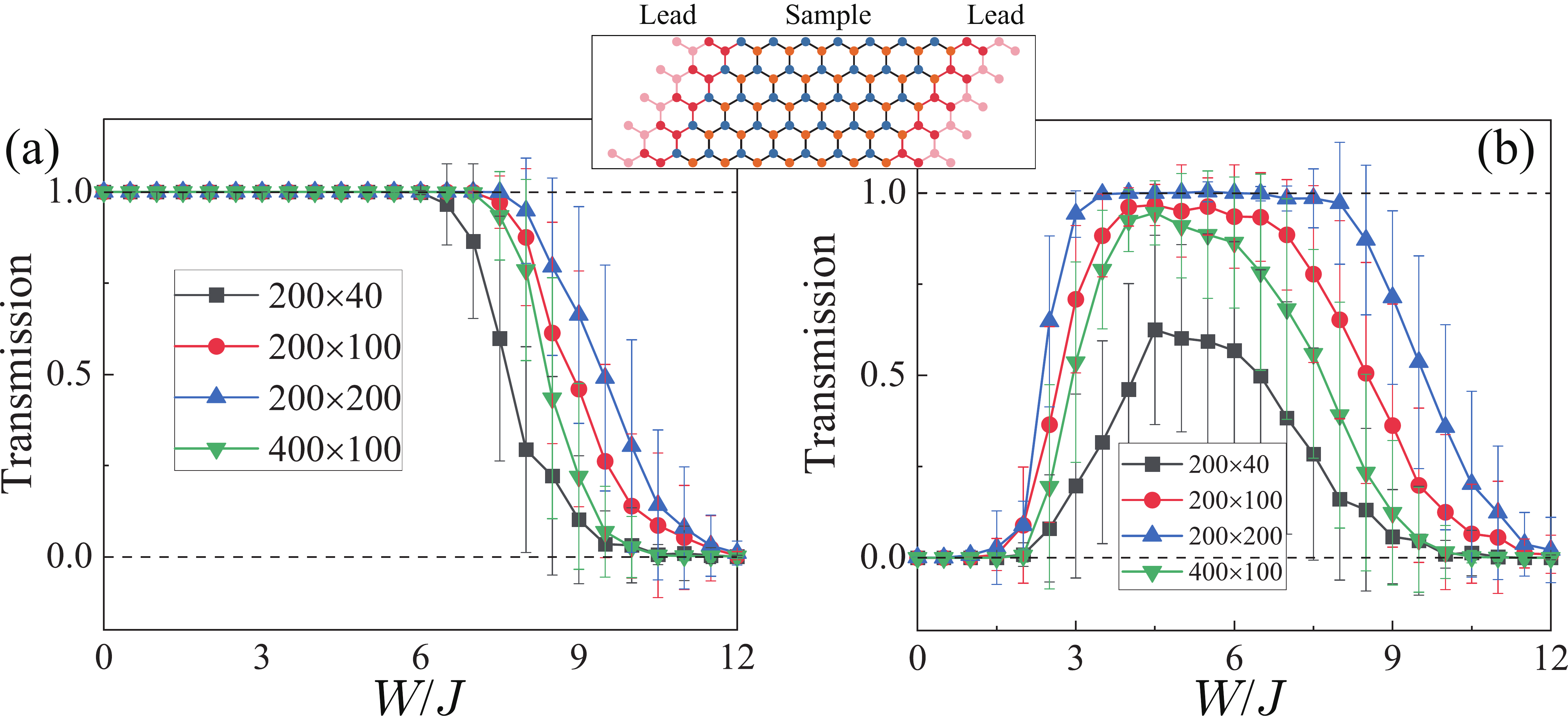}\\
\caption{Total transmission for different system sizes (length$\times$width). (a) $\varepsilon=\varepsilon_0$, $F=7J$, and $\Delta K=0$ (nontrivial in the clean limit).
(b)  $\varepsilon=\varepsilon_0$, $F=7J$, and $\Delta K=1.35J$ (trivial in the clean limit). }\label{S1}
\end{figure}

Figure~\ref{S1} shows the total transmission of $F=7J$ and $\Delta K=0$ (nontrivial in the clean limit)
and $F=7J$ and $\Delta K=1.35J$ (trivial in the clean limit) at $\varepsilon=\varepsilon_0$ for different system sizes (strip length of $L_x\times$ width $L_y$). Comparing the black squares, red circles and blue upward-pointing triangles for fixed length and different widths, the smaller the width is, the smaller the transmission is (i.e., the easier the localization is).
This phenomenon is understandable because for thinner strips, although each individual edge channel is still robust to disorder, the interscattering between the two edge channels
at two edges becomes easier. Comparing the red circles and green downward-pointing triangles for a fixed width and different lengths, the longer the strip is, the sharper the transition appears to be because
%Editor: Please ensure that the intended meaning has been maintained in the following edit.
(1) the transmission contributed by evanescent modes in longer strips is smaller than that in short strips and
(2) the interscattering between two edge channels leads to an exponential decay in the transmission.

\subsection{Transmission and the Bott index at different energies}

In the main text, we calculate the Bott index and total transmission at $\varepsilon=\varepsilon_0$.
This is the energy farthest from the bulk states, and the possible edge states are mostly localized at the edges.
Indeed, as long as $\varepsilon$ is inside the gap and not far away from $\varepsilon_0$, $\mathcal{B}_u=-\mathcal{B}_l$ are integers, and the behavior of $T(\varepsilon)$ agrees well with {$\bar{\mathcal{B}}_{u}(\varepsilon)$}. It is important to note that at the edge of topological transition, the Bott index can take non-zero or null values for different realizations of disorder, so on average, the transition is not a sharp one.

\begin{figure}[b]
  \centering
  \includegraphics[width=0.9\textwidth]{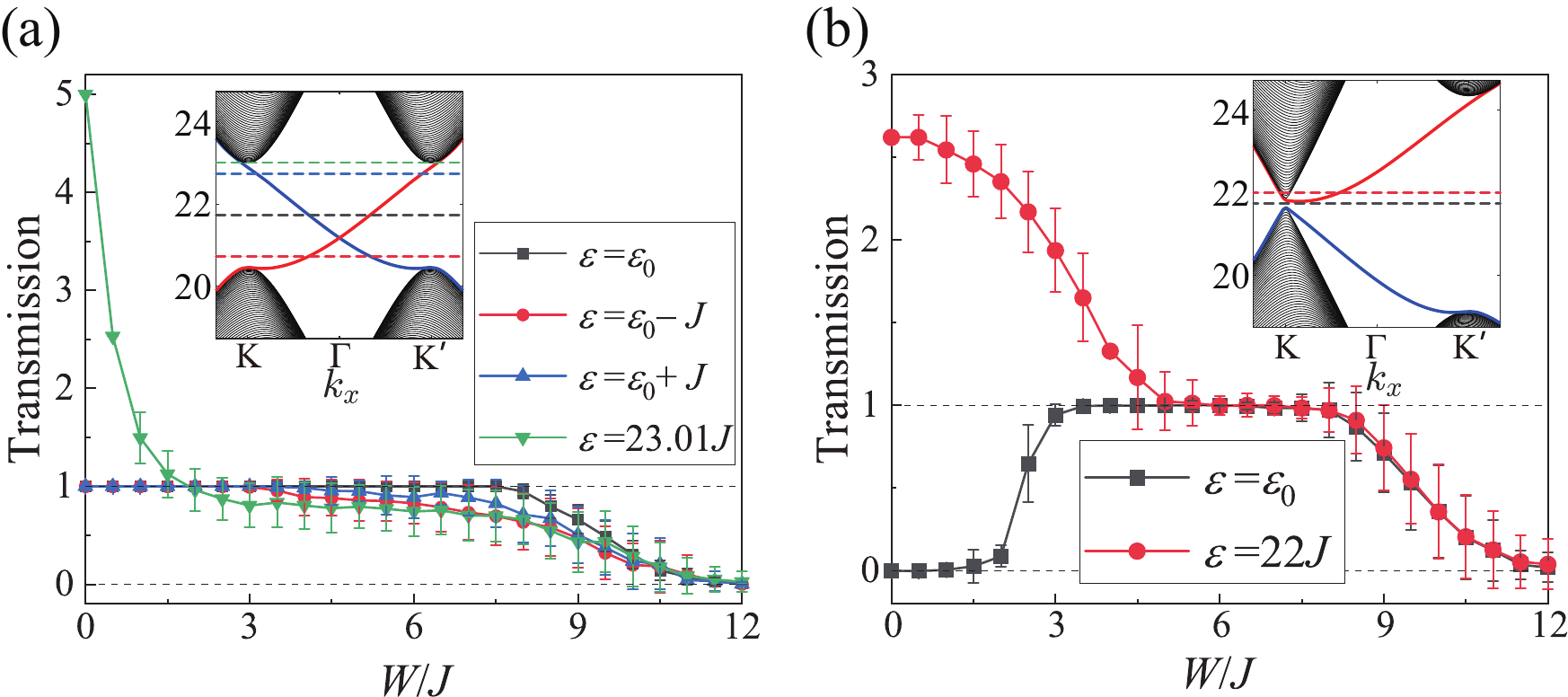}\\
\caption{Total transmission at different energies (indicated by the dashed lines in the band
structures plotted in the insets). (a) $\varepsilon=\varepsilon_0$, $F=7J$, and $\Delta K=0$ (nontrivial in the clean limit).
(b) $\varepsilon=\varepsilon_0$, $F=7J$, and $\Delta K=1.35J$ (trivial in the clean limit). }\label{S2}
\end{figure}

Figure \ref{S2} shows the total transmission versus the disorder strength at different energies.
In Fig.~\ref{S2}(a) (the nontrivial phase), when the energy is closer to the bulk bands, the edge
states are easier to scatter with the bulk modes. Thus, localization occurs more easily. When the energy is inside the bulk band
but not far away from the gap,
the transmission drops quickly from a larger value (for $\varepsilon=23.01J$, $T=5$ in the clean limit),
and the drop slows after passing 1. No plateau is observed.
In Fig.~\ref{S2}(b) (the trivial phase), when the energy is inside the bulk band but not far away
from the gap, the transmission drops from a larger value to 1 and then drops to 0 with a
$T=1$ plateau. This observation is somehow similar to that in topological Anderson insulators
(see the references in the main text).

\begin{figure}[b]
  \centering
  \includegraphics[width=0.5\textwidth]{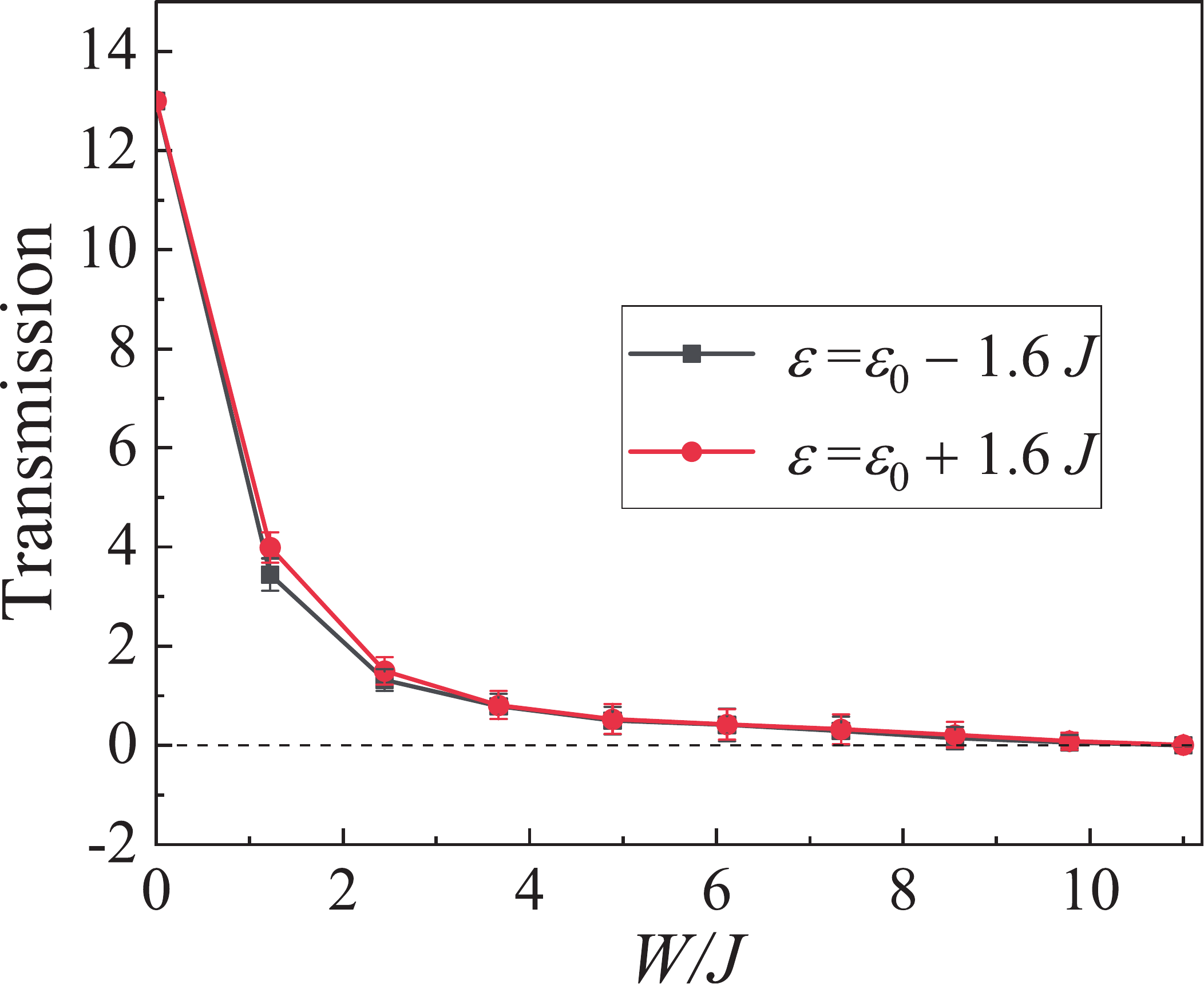}\\
\caption{Total transmission at energies inside the bulk band for $\varepsilon=\varepsilon_0$, $F=7J$, and $\Delta K=0$.}\label{S2_2}
\end{figure}

Fig.~\ref{S2_2} shows the total transmission for $\varepsilon=\varepsilon_0\pm 1.6J$, which are inside the bulk band.
At $W=0$, the total transmission $T=13$ since there are in total 13 propagating states. As the disorder increases,
the transmission decays to 0. The decay of transmission at very large disorder shows strong indication that although there is no ``band insulator" in magnonic systems, the magnons are able to be Anderson localized \cite{Martin20191}.


\begin{figure}[b]
  \centering
  \includegraphics[width=0.9\textwidth]{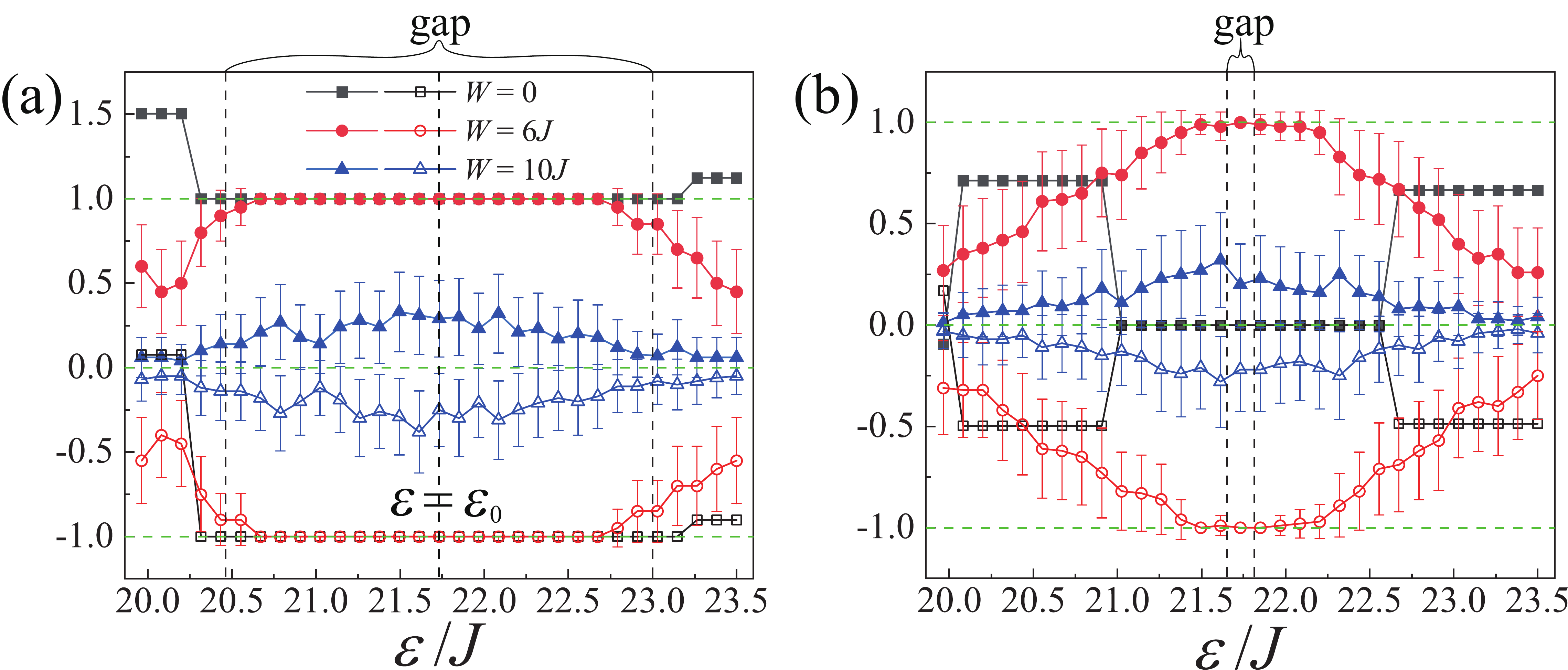}\\
\caption{Bott indices $\mathcal{B}_u(\varepsilon)$ (the Bott index of all the states above $\varepsilon$, solid symbols) and  $\mathcal{B}_l(\varepsilon)$ (the Bott index of all the states below $\varepsilon$, hollow symbols). Three disorder strengths are chosen: the clean limit ($W=0$), disordered nontrivial phase ($W=6J$), and strong-disorder localized phase ($W=10J$). (a) $\varepsilon=\varepsilon_0$, $F=7J$, and $\Delta K=0$ (nontrivial in the clean limit).
(b) $\varepsilon=\varepsilon_0$, $F=7J$, and $\Delta K=1.35J$ (trivial in clean limit). }\label{S3}
\end{figure}

Figure~\ref{S3} plots the $\varepsilon$-dependence of Bott indices $\mathcal{B}_u(\varepsilon)$ and $\mathcal{B}_l(\varepsilon)$.
Here, $\mathcal{B}_{u(l)}(\varepsilon)$ means the Bott index of all the states above (below) $\varepsilon$. In this sense, the variable $\varepsilon$ plays the role of Fermi energy
$E_F$ in electronic systems. Only when $\varepsilon$ is chosen to be inside the gap and near the gap center does the Bott index reflect the topology of the system as presented in the main text. When $\varepsilon$ is far from the gap center energy, the Bott index does not have the physical
meaning of a topological invariant. Interestingly, similar to the Bott index in the Floquet system
[Nature Communications \textbf{6}, 8335 (2015)], the clean-limit Bott indices are not integers
when $\varepsilon$ is far outside the gap. However, with disorder, the Bott indices are integers
(the noninteger value shown here is due to the average. Indeed, for each random configuration
the Bott indices are integers).

\subsection{Calculation of the self-energy}
We follow Phys. Rev. Lett. \textbf{103}, 196805 (2009) to calculate the self-energy:
\begin{equation}
\Sigma=\frac{1}{3}W^2\left(\frac{a}{2\pi}\right)^2 \int_{\text{B.Z.}}d\mathbf{k}\left[\varepsilon+i0^+-H_{\mathbf{k}}-\Sigma\right]^{-1},
\end{equation}
where $a$ is the lattice constant ($a=1$ in our case). The critical $\Delta K$ value for the topological transition (i.e., the closing gap)
is $\Delta K=\pm\frac{1}{2}\left(M-\sqrt{M^2-\frac{9}{4}F^2}\right)=1.268J$ for $F=7J$. Thus, $\Delta K=1.35J$, which we used, is in the trivial case but
is not far from the transition point. The $\Sigma^2$ component has the form
\begin{equation}
    \Sigma^2=c\left(\begin{matrix}
1 & 0 & 0 & 0 \\
0 & 1 & 0 & 0 \\
0 & 0 & -1 & 0 \\
0 & 0 & 0 & -1 \\
\end{matrix}\right),
\end{equation}
which renormalizes $\Delta K$ by $\widetilde{\Delta K}=\Delta K+c$. To drive the topological transition, the critical value of $c$ is $c=1.268J-1.35J=-0.082J$. We then scan $W$ and numerically solve for $\Sigma$ to find the critical $W=3.7J$.

\subsection{Bott index in Kagome ferromagnets}
To further justify the validity of other findings in other topological magnonic systems, we consider a 2D Kagome lattice that has been investigate in Ref. \cite{Mook20141}. For a clean $20\times20$ finite sample, we calculate the Bott indices ($\mathcal{B}$) of the three magnon bands, and compare them with the Chern numbers ($\mathcal{C}$) obtained in Figure 4 of \cite{Mook20141}. The results are listed in Table \ref{table1}.
\begin{table}[!ht]
\centering
\setlength{\tabcolsep}{7mm}{
\begin{tabular}{c|c|c|c|c|c|c}
\toprule[1pt]
$\frac{J_\text{NN}}{J_\text{N}}$ & $B_1$ & $C_1$ & $B_2$ & $C_2$ & $B_3$ & $C_3$ \\
\hline
0 & $-1$ & $-1$ & 0 & 0 & $+1$& $+1$ \\
0.5 & $-1$ & $-1$ & $+2$ & $+2$ & $-1$ & $-1$ \\
0.805 & $-3$ & $-3$ & $+4$ & $+4$ & $-1$ & $-1$ \\
0.81 & \textcolor{red}{$+3$} & $-3$ & \textcolor{red}{$-2$} & $+4$ & $-1$ & $-1$ \\
1 & $+3$ & $+3$ & $-2$ & $-2$ & $-1$ & $-1$ \\
\bottomrule[1pt]
\end{tabular}}
\caption{Comparison between Chern numbers and Bott indices for topological magnons in clean Kagome ferromagnets.
 The subscripts 1, 2, 3 label the magnon band from low energy to high energy. The parameters
and meaning of other symbols are the same as those in Ref. \cite{Mook2014}.}
\label{table1}
\end{table}
For all the four phases at $\frac{J_\text{NN}}{J_\text{N}}=0$, 0.5, 0.805, and 1, the Bott indices agree well
with the Chern numbers. The only discrepancy (label by red in the table) occurs at a very narrow phase. This may be
because the small size ($20\times20$) we used.

\end{document}